\documentclass[a4paper, 10pt, conference]{ieeeconf}      
\usepackage{FG2020}

\FGfinalcopy 

\IEEEoverridecommandlockouts                              
\usepackage{graphics} 
\usepackage{epsfig} 
\usepackage{times} 
\usepackage{amsmath} 
\usepackage{amssymb}  
\usepackage{subcaption}
\usepackage{graphicx}
\usepackage{mathtools}
\usepackage{float}

\usepackage{multirow}
\usepackage{tabularx}
\usepackage{array}

\usepackage{cite}

\def\FGPaperID{13} 

\title{\LARGE \bf
Gesture Agreement Assessment Using Description Vectors
}

\author{\parbox{16cm}{\centering
    {\large Naveen Madapana, Glebys Gonzalez and Juan Wachs}\\
    {School of Industrial Engineering, Purdue University, West Lafayette, United States.} \\
}}

\begin{document}

\ifFGfinal
\thispagestyle{empty}
\pagestyle{empty}
\else
\author{Anonymous FG2020 submission\\ Paper ID \FGPaperID \\}
\pagestyle{plain}
\fi
\maketitle

\begin{abstract}
Participatory design is a popular design technique that involves the end users in the early stages of the design process to obtain user-friendly gestural interfaces. Guessability studies followed by agreement analyses are often used to elicit and comprehend the preferences (or gestures/proposals) of the participants. Previous approaches to assess agreement, grouped the gestures into equivalence classes and ignored the integral properties that are shared between them. In this work, we represent the gestures using binary description vectors to allow them to be partially similar. In this context, we introduce a new metric referred to as soft agreement rate ($\mathcal{SAR}$) to quantify the level of consensus between the participants. In addition, we performed computational experiments to study the behavior of our partial agreement formula and mathematically show that existing agreement metrics are a special case of our approach. Our methodology was evaluated through a gesture elicitation study conducted with a group of neurosurgeons. Nevertheless, our formulation can be applied to any other user-elicitation study. Results show that the level of agreement obtained by $\mathcal{SAR}$ metric is 2.64 times higher than the existing metrics. In addition to the mostly agreed gesture, $\mathcal{SAR}$ formulation also provides the mostly agreed descriptors which can potentially help the designers to come up with a final gesture set.
\end{abstract}

\section{INTRODUCTION AND RELATED WORK}

Gestures offer an intuitive and a natural mode of interaction with computing devices. Given the ubiquity of gestural inputs in handheld smart devices and gaming consoles \cite{istance_designing_2010, wang_using_2008}, the choice of gestures plays a crucial role in the development of efficient and user-friendly gestural interfaces \cite{stern_optimal_2008}. In this paper, we utilize the principles of participatory design to obtain the best choices of gestures and further propose a new technique to measure the agreement among the participants. 
Participatory design is a popular design technique that aims to involve the end users in the early stages of the design process to obtain high acceptability interfaces \cite{vatavu_user-defined_2012, vatavu_formalizing_2015, dong2015elicitation}. It allows the stakeholders, i.e. customers or end users to provide their requirements so that the developed system complies with their preferences \cite{Muller-participatory-design, Kensing1998-participatory-design}. This technique is especially advantageous in domains requiring particular expertise i.e. radiology, neurosurgery, construction and aviation as the experts in those domains have the intrinsic knowledge about the environment which considerably affects the choice of gestures \cite{jacob_gestonurse:_2012, wachs_gestix:_2007, hettig2015exploration}. 

Guessability or user-elicitation studies are often conducted to elicit user preferences in the form of proposals for the functionalities of the interface. The nature of proposals greatly depends on the interface and its mode of operation. For instance, the proposals can be: 1. \textit{Word} choices in the case of speech-based virtual assistants such as Amazon Echo \cite{amazon-echo-wright:2016:AED:3126270}, 2. Choice of \textit{symbolic} gestures in the case of smartphones or autonomous vehicles \cite{dong2015elicitation, arefin_shimon_exploring_2016}, and 3. Choice of \textit{freehand} gestures in the case of gaming consoles \cite{ren_robust_2013, istance_designing_2010}. Our work falls under the third category in which proposals are nothing but the gestures elicited by the end users.

\begin{figure}[h]
    \centering
    \includegraphics[width = 0.7\columnwidth]{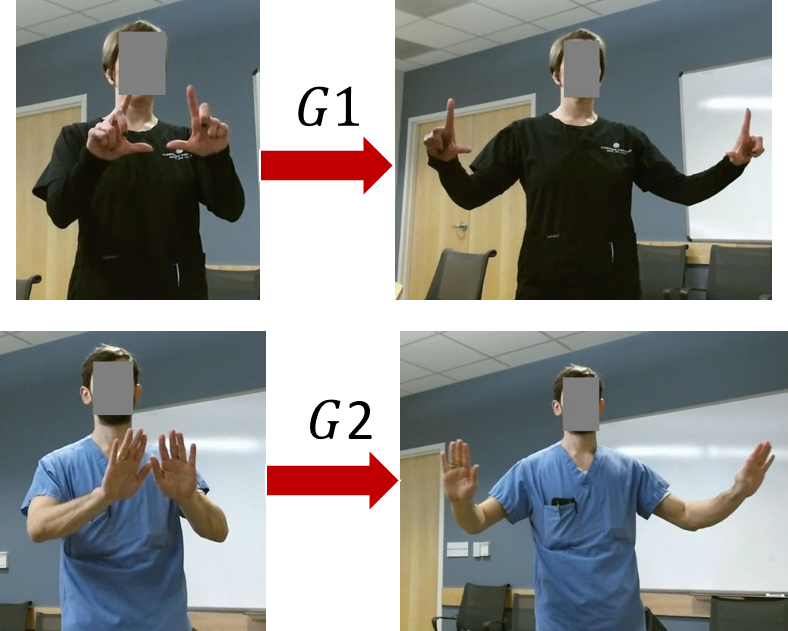}
    \caption{Gestures that are similar but not equivalent.}
    \label{fig:ex-similar-gest}
\end{figure}

Once the participants' gestures are obtained, agreement analysis is performed to quantify the agreement among users and select the best set of gestures for the final interface. The metrics introduced by Wobbrock et al. \cite{vatavu_user-defined_2012, vatavu_formalizing_2015} have been popularly used in the literature to measure the level of agreement among users. These metrics are referred to as ``agreement index'' ($\mathcal{A}_r$) \cite{vatavu_user-defined_2012} and ``agreement rate'' ($\mathcal{AR}_r$) \cite{vatavu_formalizing_2015} for command $r$. The  metric $\mathcal{A}_r$ is inconsistent at the boundary i.e. it takes a non-zero value when there is no agreement leading to an overestimation of the agreement. In this regard, Vatavu et al. proposed a new formula for agreement ($\mathcal{AR}_r$) to address this critical issue. 
Interestingly enough, Stern et al. developed a very similar formula while attempting to quantify the agreement among user-elicited proposals \cite{stern_optimal_2008}. Since their introduction, these metrics were widely adopted in several user elicitation studies to assess agreement \cite{valdes_exploring_2014, findlater2012beyond, bailly2013metamorphe, vatavu_user-defined_2012, our_plos_paper}. 

Existing methods for agreement analysis are based on the formation of equivalence groups or classes among the elicited proposals. Each class consists of proposals that are identical or very similar. Human judgement or visual inspection is often used to determine whether two proposals are identical or non-identical. Hence two proposals that are similar but not equivalent can be potentially assigned to two different equivalence classes or vice versa (refer to Fig. \ref{fig:ex-similar-gest}). In addition, the degree of similarity between the proposals is binary, i.e. a value of one refers to a pair of identical proposals and zero otherwise. Hence we refer to these methods as ``hard classification'' approaches. 

In contrast, our work allows the proposals to be partially similar, using binary description vectors. First, each proposal is decomposed into a set of atomic properties or descriptors. In the case of hand gestures, these properties could be the direction of motion, shape of the trajectory, orientation of the hand, etc. In other words, a proposal is represented as a binary vector where each element corresponds to the presence (one) or absence (zero) of a particular property. This binary vector is referred to as a soft representation or a description vector. Second, the similarity metrics that are widely used in the pattern recognition can be used to measure the extent of similarity. Next, we propose a new metric to measure the level of agreement that implicitly takes into account the description vectors. We refer to these similarity based agreement methodologies as ``soft classification'' approaches. 



The key contributions of this paper are as follows: 1. A novel methodology to perform agreement analysis for user-elicitation studies by incorporating the description vectors, 2. A mathematical proof showing that the existing agreement metrics are a special case of our approach. 3. An empirical relationship between the agreement metric and the average percentage of participants that agreed on a particular gesture.


\section{Methodology}


Let us start by defining the notations. Let $C$ be the number of commands or referents, $d$ be the number of descriptors, and subscript $r$ denote the command $r$. Let $P_r$ be the set of all proposals or gestures, where $r = 1, \dots, C$. Let $u_r \leq \lvert P_r \rvert$ be the number of unique gestures. Let $P^i_r$ be a subset of proposals that are considered identical. Thus, $\lvert P^i_r \rvert$ would be the number of identical proposals in the set $i$. We use $|.|$ to denote the number of elements in a set and use $||.||$ to denote the L2 norm of a vector. Let $i^{th}$ proposal be represented as a binary description vector $S^i_r \in \{0, 1\}^d$, where $i = 1, 2, \dots, N_r$. The total number of proposals ($N_r$) for the command $r$ can be represented as the following (Eq. \ref{eq:n_r}). Note that all referents have same number of proposals.

\begin{equation} \label{eq:n_r}
N = N_r = \sum_{i=1}^{u_r} \lvert P^i_r \rvert = \lvert P_r \rvert
\end{equation}

The most commonly used  agreement metric defined by Wobbrock et al. 
\cite{vatavu_formalizing_2015} is given in equation \ref{eq:wob-2015}. $\mathcal{AR}_r$ is the level of agreement for command $r$. 


\begin{equation} \label{eq:wob-2015}
\mathcal{AR}_r = \frac{1}{N(N -1 )} \sum_{i=1}^{u_r}\lvert P^i_r \rvert (\lvert P^i_r \rvert -1)  
\end{equation}


\subsection{Soft Agreement Rate (SAR)}

To define an agreement index, we need a similarity metric that measures closeness between two vectors (binary in our case). Popular similarity metrics include but not limited to cosine \cite{nguyen_cosine_2010}, Jaccard \cite{niwattanakul_using_2013} and Hamming similarity \cite{bayardo_scaling_2007}, which vary between zero and one (one when the vectors are equivalent and zero when they are orthogonal). Note that the Jaccard similarity does not consider zero-zero (a descriptor being absent in both the gestures)  as an agreement but only considers one-one as an agreement. It complies with the context of agreement analysis because when two gestures lack a descriptor, it does not necessarily imply an agreement. Given the sparsity in the description vectors, the cosine distance is a good alternative to measure the similarity \cite{bayardo_scaling_2007}. Hamming similarity can not be used to measure similarity as it considers zero-zero as an agreement.

We propose an agreement metric referred to as Soft Agreement Rate (\textit{SAR}). $\mathcal{SAR}_r$ is the agreement for command $r$ which is defined as a mean of the Jaccard similarity $J$ applied to all possible pairwise combinations of binary vectors corresponding to the proposals in $P_r$ (Eq. \ref{eq:sar-r}). An overall agreement rate $(\mathcal{SAR})$ is defined as a mean of $\mathcal{SAR}_r$ for all referents $r$. The mathematical representation of $\mathcal{SAR}$ relates to $\mathcal{AR}$ in terms of considering all possible pairwise combinations. Similar to the $\mathcal{AR}$, the proposed metric $\mathcal{SAR}$ takes a value of $0$ where there is no agreement and takes a value of $1$ when all participants agree on a proposal. \\

\begin{equation} \label{eq:sar-r}
\mathcal{SAR}_{r} = \frac{2}{N \left( N -1 \right)} \sum_{j=k+1}^{N} \sum_{k=1}^{N}J\left(  S ^j_r, S^k_r\right)
\end{equation}


Where, $S^j_r$ and $S^k_r$ represent the binary vectors of $j^{th}$ and $k^{th}$ proposal. Since the Jaccard similarity between two zero vectors is not defined, we propose a conditional definition for this similarity metric (Eq. \ref{eq:cond-jac}). Our methodology can be easily generalized to the cosine similarity.

\begin{equation} \label{eq:cond-jac}
    J(a,b)= 
\begin{dcases}
    0,& \text{if } \left\Vert a\right\Vert + \left\Vert b\right\Vert = 0 \\
    \frac{a.b}{\left\Vert a\right\Vert^2 + \left\Vert b\right\Vert^2 -a.b},              & \text{otherwise}
\end{dcases}
\end{equation}


Where $a$ and $b$ are the binary vectors, and $a.b$ denotes the dot product between the vectors $a$ and $b$. 

\subsection{Relation to Existing Metrics}

In this section, we prove that the metric $\mathcal{AR}$ is a special case of our metric $\mathcal{SAR}$. Let $\mathcal{SAR}^{hard}$ denote the soft agreement rate when the gestures are treated as rigid entities and are grouped into equivalence classes instead of as a combination of descriptors. In machine learning, it is common to represent a distinct equivalence class (gestures in this case) as a one hot (OH) vector.

There are $u_r$ distinct equivalence classes or unique gestures for referent $r$. In this special case, each unique gesture is assigned to a distinct OH vector of length $u_r$. This implies that all the gestures in the set $P^i_r$ are assigned to the same OH vector. The Jaccard similarity between two identical OH vectors is unity and two distinct OH vectors is 0. This nullifies all of the distinct pairwise OH vectors. The resulting nonzero combinations are obtained from the pairwise combinations within the subset $P^i_r$. That is,  $J(S^j_r,S^k_r; j \neq k) = 0$ and $(S^j_r,S^k_r; j = k) = 1$.

\begin{equation} \nonumber
\Rightarrow{} \sum_{j=k+1}^{N} \sum_{k=1}^{N}J(S^j_r,S^k_r) =  \frac{1}{2}\sum_{i=1}^{u_r} \lvert P^i_r \rvert (\lvert P^i_r \rvert -1)  
\end{equation}


By substituting the above in Eq. \ref{eq:sar-r}, we will get, $\mathcal{SAR}^{hard}_r = \mathcal{AR}_{r}$. Interestingly enough, the resulting equation is exactly equal to the one proposed by Vatavu et al. \cite{vatavu_formalizing_2015}. This proves that our approach is general enough to adapt for both soft and hard representation of gestures. 

\subsection{Interpretation of Agreement Index}
We know that the agreement rate is related to the number of participants that agreed on a particular proposal. Nevertheless, there is no direct theoretical / empirical relation between the agreement value and the average percentage of participants that agreed on a proposal ($\eta$). For example, a value of $\eta = 0.40$ indicates that $40\%$ of participants agreed on a proposal. In this section, we present an elegant way of interpreting the level of agreement. Given the agreement rate, the main idea is to estimate the average number of subjects that agreed on a particular proposal. 

For instance, assume that we have 10 proposals for a referent $r$ and consider the following two scenarios: a) Four proposals are equivalent and the rest are different ($\mathcal{AR} = 0.13, \eta = 0.40$) and b) Six proposals are equivalent ($\mathcal{AR} = 0.33, \eta = 0.60$) and the rest different. In each of these cases, it is unclear how the $\mathcal{AR}$ is related to $\eta$. Note that $\eta ^2 \approx \mathcal{AR}$ in these simplistic scenarios. In more complex scenarios such as when 4 out of 10 subjects picked a particular gesture, 3 out of 10 picked another gesture, and rest of subjects picked different gestures, the value of $\eta$ can not be quantified.

Thus, we propose an empirical relation between $\eta$ and level of agreement which is a numerical approximation of $\eta$ in terms of agreement value. Given the quadratic nature (pairwise combinations) of the agreement metrics, we propose that value of $\eta$ is approximately equal to the square root of the $\mathcal{SAR}$ or $\mathcal{AR}$ (Eq. \ref{eq:sqrt-ar} and \ref{eq:sqrt-sar}). In the case of hard representations, $\eta$ indicates the average number of subjects that agreed on a gesture. However, for soft representations, $\eta$ gives the average number of subjects that agreed on a set of gesture descriptors. Though $\eta$ is an empirical and an approximate measure of an average percentage, it is meant to provide a qualitative interpretation of the agreement values.

\begin{equation} \label{eq:sqrt-ar}
\eta^{\mathcal{AR}}_r \approx \sqrt{\mathcal{AR}_{r}}
\end{equation}

\begin{equation} \label{eq:sqrt-sar}
\eta^{\mathcal{SAR}}_r \approx \sqrt{\mathcal{SAR}_{r}}
\end{equation}

\section{Experiments and Results}
\subsection{Case Study}
\begin{figure}[h]
  \centering
  \includegraphics[width=0.9\columnwidth]{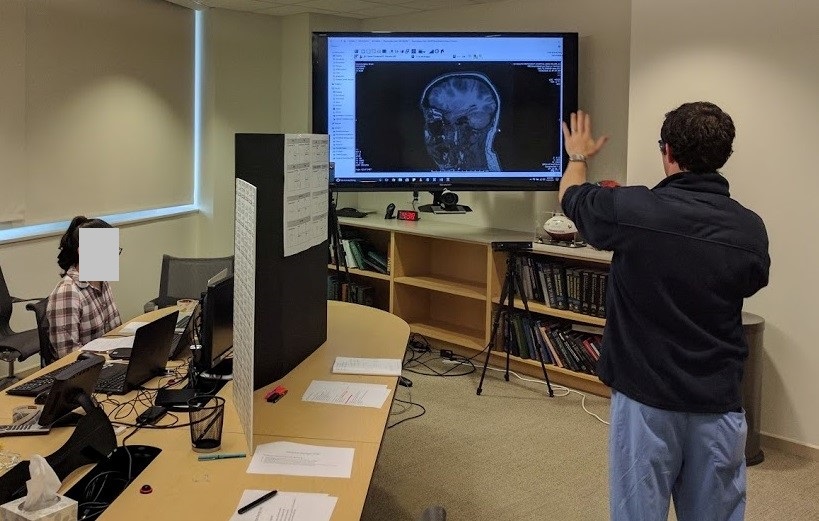}
  \caption{Gesture elicitation study with neurosurgeons.}
    ~\label{fig:surgeon-gestures}
\end{figure}

Our approach was evaluated and tested using the gestures elicited by the neurosurgeons (refer to Fig. \ref{fig:surgeon-gestures}) in a guessability study conducted by Madapana et al. \cite{our_plos_paper}. In their study, a group of nine neurosurgeons (participants) were asked to create gestures (proposals) for each of the 28 commands (referents) present in a radiology image browser. A total of 252 gestures were considered. Next, a subset of 55 gesture descriptors proposed in the literature \cite{madapana_database_2019, lascarides_formal_2009} were used in this paper as shown in Table \ref{tab:sds} in order to create the soft representations. In other words, each of the 252 gestures were represented as a 55 dimensional binary description vectors. 

Two experiments were conducted in order to compute the level of agreement using: $\mathcal{AR}$ and $\mathcal{SAR}$. In the first experiment, a group of six participants were asked to group the gestures for each referent into equivalent classes. Hence each equivalent class consists of gestures that are physically similar. This procedure was repeated for all of the 28 referents. 
We used these groupings to compute the level of agreement using $\mathcal{AR}$ formulation (refer to Equation \ref{eq:wob-2015}) as shown in table \ref{tab:tab1}.


\begin{center}
\begin{table}[H]
\centering
  \caption{A list of gesture descriptors. Note that L, R, U, D, F and B stand for left, right, up, down, forward and backward. CW and CCW stand for clockwise and counter clockwise.}
  \label{tab:sds}
\begin{tabularx}{\linewidth}{|>{\hsize=.1\hsize}X|>{\hsize=1.4\hsize}X|>{\hsize=1.5\hsize}X|}
    \hline
    \centering \textbf{\#} & \multicolumn{2}{c|}{\textbf{Movement descriptors}}\\
    \hline \hline
    \centering 1 & Dominant hand &  \multirow{2}{*}{\parbox[t][0.8cm]{3.6cm}{ L, R, U, D, F, B, CW, CCW, Iterative, and Circular}}\\
    \cline{1-2}
    \centering 2 & Non-Dominant hand & \\
    \cline{1-3}
   \centering 3 & Overall hand motion & Inward, Outward flow, Circular\\
    \hline
    \hline
    \centering  & \multicolumn{2}{c|}{\textbf{Orientation descriptors}}\\
    \hline
    \hline
    \centering 4 & Dominant hand &  \multirow{2}{*}{\parbox[c][0.6cm]{3.6cm}{Left, Right, Up, Down, Forward and Backward}}\\
    \cline{1-2}
    \centering 5 & Non-Dominant hand & \\
    \hline
    \hline
    \centering  & \multicolumn{2}{c|}{\textbf{State of the hand }}\\
    \hline
    \hline    
    \centering 6 & Dominant hand & \multirow{2}{*}{\parbox[c][0.6cm]{3.7cm}{C shape, V shape, Fist, 1, 2, 3, 4 open fingers, open palm}}\\
    \cline{1-2}
    \centering 7 & Non-Dominant hand & \\
    \hline
    \hline
    \centering & \multicolumn{2}{c|}{\textbf{Other descriptors }}\\
    \hline
    \hline    
    \centering 8 & Avg. position of gesture & Above, Medium and Low\\
    \hline
\end{tabularx}
\end{table}
\end{center}

\vspace{-2.6em}

In the second experiment, the same group of six participants were asked to annotate each of the 252 gestures with respect to their gesture descriptors. We developed a web interface that facilitates annotating descriptors for each gesture. This interface consisted of a window that plays the gesture video and a set of questions asking participants to annotate whether a particular descriptor is present or not. The annotations were automatically parsed to generate the description vector. We used these annotations to compute agreement using $\mathcal{SAR}$ metric as shown in table \ref{tab:tab1}.\\



\subsection{Distribution of $\mathcal{SAR}$ metric}
\begin{figure}[h]
    \centering
    \includegraphics[width=0.5\textwidth]{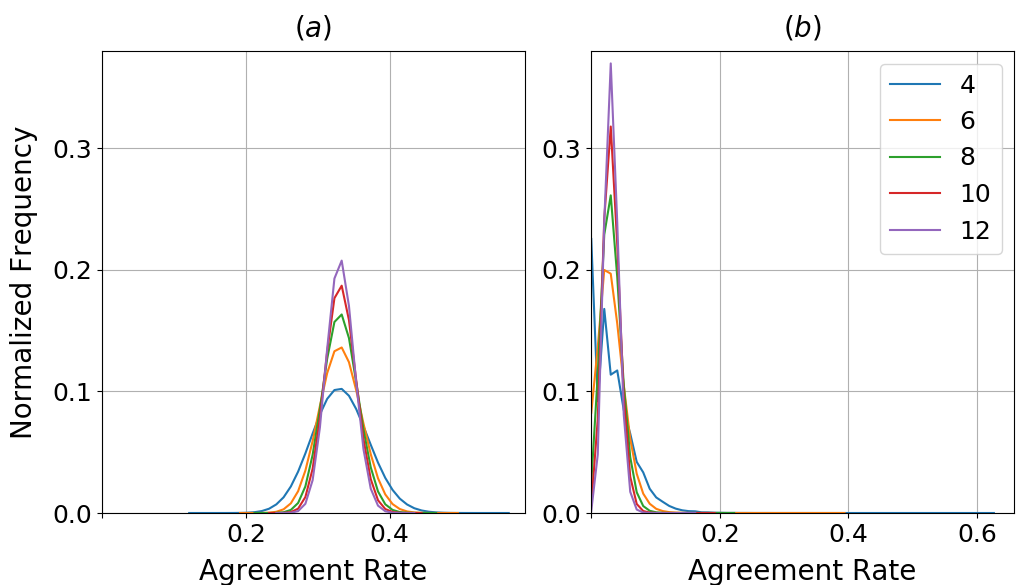}
    \caption{Probability distribution of $\mathcal{SAR}$ with varying number of subjects and $d = 55$ when the input gesture description data is sampled from a Bernoulli distribution with p(1) = 0.50 (left) and 0.93 (right).}
    \label{fig:pdf}
\end{figure}

This section presents the probability distribution function $\mathcal{PDF}(S | d = 55)$ of the $\mathcal{SAR}$ metric by varying the number of subjects ($S$). This involved forming binary gesture description vectors of dimension $d = 55$ for each of the gesture proposals elicited by the $S$ subjects. These description vectors were sampled from a Bernoulli distribution with probability $P(1) = 0.5$. First, the random descriptions were generated and then, the level of agreement using $\mathcal{SAR}$ was computed. This procedure is repeated for $10^7$ iterations and the normalized histogram of agreement values was constructed using 100 bins of equal intervals. Fig. \ref{fig:pdf}a shows the PDF of $\mathcal{SAR}$ when the no. of subjects vary i.e. $\mathcal{PDF}(S | d = 55)$. The shape of the distribution resembles the bell curve with peak occurring at 0.33 approximately. For $S = 9$ and $D = 55$, the cumulative probability $P(\mathcal{SAR} \leq 0.35) = 0.88$ while $P(\mathcal{SAR} \leq 0.40) = 0.999$. 

\begin{table}[h]
\large
\caption{Values of agreement ($\mathcal{AR}$ and $\mathcal{SAR}$) and estimated average number of subjects that agreed on particular command or a set of descriptors ($\eta^{AR}$ and $\eta^{SAR}$).} 
\label{tab:tab1}
\resizebox{\columnwidth}{!}
{%
\begin{tabular}{|c|c|c|c|c|}
    \hline 
    \textbf{Command} & $\mathcal{AR}$ & $\eta^{\mathcal{AR}}$ (\%) & $\mathcal{SAR}$ & $\eta^{\mathcal{SAR}}$ (\%) \\
    \hline \hline
    Scroll up         & $0.08$ & $28.87$ & $0.30$ &  $54.50$ \\
    \hline
    Flip horizontal   & $0.13$ & $36.51$ & $0.39$ &  $62.66$ \\
    \hline
    Rotate CW         & $0.20$ & $44.72$ & $0.32$ &  $56.76$ \\
    \hline
    Zoom in           & $0.19$ & $44.10$ & $0.22$ &  $46.51$ \\
    \hline
    Zoom out          & $0.23$ & $47.73$ & $0.22$ &  $47.08$ \\
    \hline
    Panel left        & $0.15$ & $38.73$ & $0.21$ &  $45.79$ \\
    \hline
    Pan left          & $0.13$ & $36.51$ & $0.29$ &  $53.47$ \\
    \hline
    Ruler measure     & $0.12$ & $34.96$ & $0.19$ &  $43.90$ \\
    \hline
    Window open       & $0.07$ & $25.82$ & $0.23$ &  $47.73$ \\
    \hline
    Inc. contrast     & $0.06$ & $24.72$ & $0.36$ &  $59.97$ \\
    \hline
    Layout           & $0.09$ & $30.73$ & $0.37$ &  $60.47$ \\
    \hline
    Preset           & $0.07$ & $25.82$ & $0.32$ &  $56.99$ \\
    \hline
    \hline

  \textbf{Mean} $\pm$ \textbf{Std} & $0.11 \pm 0.10$ & $33.7 \pm 7.41$ & $0.29 \pm 0.05$ & $54 \pm 5.0$ \\
     \hline
  \end{tabular}
}
\end{table}


Note that these PDFs were constructed assuming that the input data resembles a Bernoulli distribution with $P(1) = 0.5$. However, the actual gesture description data is sparse with zeroes occurring more frequently than ones $P(1) = 0.07$. Hence we conducted a new set of similar computational experiments to construct the PDF of $\mathcal{SAR}$ when we feed the data that resembles the actual data (Bernoulli distribution with $P(1) = 0.07$). Fig. \ref{fig:pdf}b shows the PDF of our metric when the parameter $S$ is varied. Note that the shape of this distribution does not look like a bell curve anymore and the peak occurs between 0.0 and 0.1. For $S = 9$, the cumulative probability $P(\mathcal{SAR} \leq 0.04) = 0.84$ while $P(\mathcal{SAR} \leq 0.07) = 0.99$.

\subsection{Discussion}
$\mathcal{SAR}$ formulation considers the properties of the gestures and hence produces relatively higher agreement rates in comparison to previous approaches. These results are expected, given that the $\mathcal{SAR}$ formulation considers the partial similarity between the gestures. This indicates that participants are more likely to agree on high-level properties of the gestures than agreeing on the gesture itself. 
It clearly indicates that the soft representations provide deeper understanding about the properties that the domain experts agree on, even when the gestures are not identical. Indirectly, the $\mathcal{SAR}$ metric focuses on what experts emphasize in the gestures rather than on their plain spatio-temporal appearance.


Our formulation is particularly advantageous when the values of $\mathcal{AR}$ are very low i.e. $\mathcal{AR} < 0.1$ \cite{vatavu_formalizing_2015}. Given a very little agreement, it is very hard to determine the final gesture as the mostly agreed gesture is chosen by very few participants. In such cases, the system interface designers will be greatly benefited from $\mathcal{SAR}$ as it allows them to determine the properties of the final gesture. Similarly, when the difference between $\mathcal{SAR}$ and $\mathcal{AR}$ is large as in the case of \textit{scroll up}, the interface designers are recommended to utilize both the methodologies when determining the final gesture lexicon.

\section{Conclusion}
Existing techniques for agreement analysis strongly rely on the formation of equivalence classes and ignore the integral properties that are shared between the gestures. In this work, we represent gestures as binary description vectors by decomposing the gestures into a set of measurable properties. Next, we propose an agreement metric referred to as Soft Agreement Rate ($\mathcal{SAR}$) that incorporates the description vectors into the agreement analysis. We further show that the existing agreement metrics are a special case of our approach when the equivalence classes themselves act as descriptors. Moreover, we propose an empirical relation between the level of agreement and the average number of participants that agreed on a particular gesture or a set of descriptors. We evaluated our approach using a gesture elicitation study conducted with a group of neurosurgeons. Our results show that $\mathcal{SAR}$ metric yields considerably higher agreement rates than the existing metrics. In addition to providing the mostly agreed gesture, $\mathcal{SAR}$ formulation also provides the mostly agreed descriptors. This can potentially help the interface designers to choose the final gesture when the level of agreement is considerably low.  




\bibliography{fg2020}

\begin{thebibliography}{10}

\bibitem{istance_designing_2010}
H.~Istance, A.~Hyrskykari, L.~Immonen, S.~Mansikkamaa, and S.~Vickers,
  ``Designing {Gaze} {Gestures} for {Gaming}: {An} {Investigation} of
  {Performance},'' in {\em Proceedings of the 2010 {Symposium} on
  {Eye}-{Tracking} {Research} \& {Applications}}, {ETRA} '10, (New York, NY,
  USA), pp.~323--330, ACM, 2010.

\bibitem{wang_using_2008}
Y.~Wang, T.~Yu, L.~Shi, and Z.~Li, ``Using human body gestures as inputs for
  gaming via depth analysis,'' in {\em 2008 {IEEE} {International} {Conference}
  on {Multimedia} and {Expo}}, pp.~993--996, June 2008.

\bibitem{stern_optimal_2008}
H.~I. Stern, J.~P. Wachs, and Y.~Edan, ``Optimal {Consensus} {Intuitive} {Hand}
  {Gesture} {Vocabulary} {Design},'' in {\em 2008 {IEEE} {International}
  {Conference} on {Semantic} {Computing}}, pp.~96--103, Aug. 2008.

\bibitem{vatavu_user-defined_2012}
R.-D. Vatavu, ``User-defined {Gestures} for {Free}-hand {TV} {Control},'' in
  {\em Proceedings of the 10th {European} {Conference} on {Interactive} {TV}
  and {Video}}, {EuroITV} '12, (New York, NY, USA), pp.~45--48, ACM, 2012.

\bibitem{vatavu_formalizing_2015}
R.-D. Vatavu and J.~O. Wobbrock, ``Formalizing {Agreement} {Analysis} for
  {Elicitation} {Studies}: {New} {Measures}, {Significance} {Test}, and
  {Toolkit},'' in {\em Proceedings of the 33rd {Annual} {ACM} {Conference} on
  {Human} {Factors} in {Computing} {Systems}}, {CHI} '15, (New York, NY, USA),
  pp.~1325--1334, ACM, 2015.

\bibitem{dong2015elicitation}
H.~Dong, A.~Danesh, N.~Figueroa, and A.~El~Saddik, ``An elicitation study on
  gesture preferences and memorability toward a practical hand-gesture
  vocabulary for smart televisions,'' {\em IEEE Access}, vol.~3, pp.~543--555,
  2015.

\bibitem{Muller-participatory-design}
M.~J. Muller and S.~Kuhn, ``Participatory design,'' {\em Commun. ACM}, vol.~36,
  pp.~24--28, June 1993.

\bibitem{Kensing1998-participatory-design}
F.~Kensing and J.~Blomberg, ``Participatory design: Issues and concerns,'' {\em
  Computer Supported Cooperative Work (CSCW)}, vol.~7, pp.~167--185, Sep 1998.

\bibitem{jacob_gestonurse:_2012}
M.~G. Jacob, Y.~T. Li, and J.~P. Wachs, ``Gestonurse: {A} multimodal robotic
  scrub nurse,'' in {\em 2012 7th {ACM}/{IEEE} {International} {Conference} on
  {Human}-{Robot} {Interaction} ({HRI})}, pp.~153--154, Mar. 2012.

\bibitem{wachs_gestix:_2007}
J.~Wachs, H.~Stern, Y.~Edan, M.~Gillam, C.~Feied, M.~Smith, and J.~Handler,
  ``Gestix: {A} {Doctor}-{Computer} {Sterile} {Gesture} {Interface} for
  {Dynamic} {Environments},'' in {\em Soft {Computing} in {Industrial}
  {Applications}}, pp.~30--39, Springer, Berlin, Heidelberg, 2007.
\newblock DOI: 10.1007/978-3-540-70706-6\_3.

\bibitem{hettig2015exploration}
J.~Hettig, A.~Mewes, O.~Riabikin, M.~Skalej, B.~Preim, and C.~Hansen,
  ``Exploration of 3d medical image data for interventional radiology using
  myoelectric gesture control,'' in {\em Proceedings of the Eurographics
  Workshop on Visual Computing for Biology and Medicine}, pp.~177--185,
  Eurographics Association, 2015.

\bibitem{amazon-echo-wright:2016:AED:3126270}
S.~Wright, {\em Amazon Echo Dot: Amazon Dot Advanced User Guide (2017 Updated)
  Step-by-Step Instructions to Enrich Your Smart Life! (Amazon Echo, Dot, Echo
  Dot, Amazon Echo User Manual, Echo Dot Ebook, Amazon Dot)}.
\newblock USA: CreateSpace Independent Publishing Platform, 2016.

\bibitem{arefin_shimon_exploring_2016}
S.~S. Arefin~Shimon, C.~Lutton, Z.~Xu, S.~Morrison-Smith, C.~Boucher, and
  J.~Ruiz, ``Exploring {Non}-touchscreen {Gestures} for {Smartwatches},'' in
  {\em Proceedings of the 2016 {CHI} {Conference} on {Human} {Factors} in
  {Computing} {Systems}}, {CHI} '16, (New York, NY, USA), pp.~3822--3833, ACM,
  2016.

\bibitem{ren_robust_2013}
Z.~Ren, J.~Yuan, J.~Meng, and Z.~Zhang, ``Robust {Part}-{Based} {Hand}
  {Gesture} {Recognition} {Using} {Kinect} {Sensor},'' {\em IEEE Transactions
  on Multimedia}, vol.~15, pp.~1110--1120, Aug. 2013.

\bibitem{valdes_exploring_2014}
C.~Valdes, D.~Eastman, C.~Grote, S.~Thatte, O.~Shaer, A.~Mazalek, B.~Ullmer,
  and M.~K. Konkel, ``Exploring the {Design} {Space} of {Gestural}
  {Interaction} with {Active} {Tokens} {Through} {User}-defined {Gestures},''
  in {\em Proceedings of the 32Nd {Annual} {ACM} {Conference} on {Human}
  {Factors} in {Computing} {Systems}}, {CHI} '14, (New York, NY, USA),
  pp.~4107--4116, ACM, 2014.

\bibitem{findlater2012beyond}
L.~Findlater, B.~Lee, and J.~Wobbrock, ``Beyond qwerty: augmenting touch screen
  keyboards with multi-touch gestures for non-alphanumeric input,'' in {\em
  Proceedings of the SIGCHI Conference on Human Factors in Computing Systems},
  pp.~2679--2682, ACM, 2012.

\bibitem{bailly2013metamorphe}
G.~Bailly, T.~Pietrzak, J.~Deber, and D.~J. Wigdor, ``M{\'e}tamorphe:
  augmenting hotkey usage with actuated keys,'' in {\em Proceedings of the
  SIGCHI Conference on Human Factors in Computing Systems}, pp.~563--572, ACM,
  2013.

\bibitem{our_plos_paper}
N.~Madapana, G.~Gonzalez, R.~Rodgers, L.~Zhang, and J.~P. Wachs, ``Gestures for
  picture archiving and communication systems (pacs) operation in the operating
  room: Is there any standard?,'' {\em PLOS ONE}, vol.~13, pp.~1--13, 06 2018.

\bibitem{nguyen_cosine_2010}
H.~V. Nguyen and L.~Bai, ``Cosine {Similarity} {Metric} {Learning} for {Face}
  {Verification},'' in {\em Computer {Vision} – {ACCV} 2010}, Lecture {Notes}
  in {Computer} {Science}, pp.~709--720, Springer, Berlin, Heidelberg, Nov.
  2010.

\bibitem{niwattanakul_using_2013}
S.~Niwattanakul, J.~Singthongchai, E.~Naenudorn, and S.~Wanapu, ``Using of
  {Jaccard} coefficient for keywords similarity,'' in {\em Proceedings of the
  {International} {MultiConference} of {Engineers} and {Computer}
  {Scientists}}, vol.~1, 2013.

\bibitem{bayardo_scaling_2007}
R.~J. Bayardo, Y.~Ma, and R.~Srikant, ``Scaling {Up} {All} {Pairs} {Similarity}
  {Search},'' in {\em Proceedings of the 16th {International} {Conference} on
  {World} {Wide} {Web}}, {WWW} '07, (New York, NY, USA), pp.~131--140, ACM,
  2007.

\bibitem{madapana_database_2019}
N.~{Madapana} and J.~{Wachs}, ``Database of gesture attributes: Zero shot
  learning for gesture recognition,'' in {\em 2019 14th IEEE International
  Conference on Automatic Face Gesture Recognition (FG 2019)}, pp.~1--8, May
  2019.

\bibitem{lascarides_formal_2009}
A.~Lascarides and M.~Stone, ``A {Formal} {Semantic} {Analysis} of {Gesture},''
  {\em Journal of Semantics}, vol.~26, no.~4, p.~393, 2009.

\end{thebibliography}
\bibliographystyle{ieeetr}

\end{document}